\documentclass[12pt]{iopart}

\usepackage{graphicx,amsfonts,amssymb}

\begin{document}
\title[PC Preconditioners for  NK Solvers in Fluid Problems]{Predictor-Corrector Preconditioners for  Newton-Krylov Solvers in Fluid Problems}
\author{Giovanni Lapenta}

\address{Departement Wiskunde, Katholieke Universiteit Leuven, Celestijnenlaan 200B, 3001 Leuven, Belgium}
\ead{giovanni.lapenta@wis.kuleven.be}

\author{S. Ju}
\address{University of New Mexico, NM, USA}

\maketitle

\begin{abstract}
We propose an alternative implementation of preconditioning
techniques for the solution of non-linear problems. Within the
framework of Newton-Krylov methods, preconditioning techniques are
needed to improve the performance of the solvers. We propose a
different implementation approach to re-utilize existing
semi-implicit methods to precondition fully implicit non-linear
schemes. We propose a predictor-corrector approach where the fully
non-linear scheme is the corrector and the pre-existing
semi-implicit scheme is the predictor. The advantage of the proposed
approach is that it allows to retrofit existing codes, with only
minor modifications, in particular avoiding the need to reformulate
existing methods in terms of variations, as required instead by
other approaches now currently used. To test the performance of the
approach we consider a non-linear diffusion problem and the standard
driven cavity problem for incompressible flows.
\end{abstract}

\section{Introduction}
A classic problem of computational science and engineering is the
search for an efficient numerical scheme for solving non-linear
time-dependent partial differential equations. Explicit and
semi-implicit methods can provide simple solution techniques but are
seriously limited by time step limitations for stability (explicit
methods) and accuracy (semi-implicit methods).

Recently, significant progress has been made in the development of
fully implicit approaches for solving nonlinear problems: the
 Newton-Krylov (NK) method~\cite{kelley,saad-nlkrylov}. The method is developed
from the Newton iterative method, by applying a  linear iterative
solver to the Jacobian equation for the Newton step and terminating
that iteration when a suitable convergence criterion
holds.

For the solution of the linear Jacobian equation, Krylov methods are
often the choice, leading to the Newton-Krylov (NK) approach.
However, for most cases, Krylov solvers can be extremely
inefficient. The need for good preconditioners techniques becomes a
constraining factor in the development of NK solvers~\cite{saad-book}.

In a number of fields, recent work based on  multi-grid and
physics-based preconditioners~\cite{review,knoll,pernice} have demonstrated
extremely competitive performances.

In the present study, we present a different implementation of
preconditioning: the predictor-corrector (PC) preconditioner. The
approach has two novelties. First, it preconditions directly the
non-linear equations rather than the linear Jacobian equation for
the Newton step. The idea is not new~\cite{kelley}, but it is
implemented here in a new way that leads to great simplifications of
the implementation. We note that this simplification is designed
also to minimize the effort in refitting existing semi-implicit
codes into full fledged implicit codes, representing perhaps a
greater advance in software engineering than in computational
science. Second, we test new ways of preconditioning the equations
by using a combination of predictor-corrector semi-implicit
preconditioning.

The fundamental idea is to use a predictor to advance a
semi-implicit discretization of the governing equations and use a
corrector Newton step to correct for the initial state of the
predictor step. The typical NK solver is used to compute the unknown
value of the state vector at the end of the time step ${\bf x}^{1}$
from its known value at the previous time step ${\bf x}^0$. Instead,
we use the Newton method to iterate for a modification of the actual
known state $ {\bf x}^{*}$ from the previous time step to find a
modified {\it "previous"} state that makes the semi-implicit
predictor step give the solution of the fully implicit method.

Two advantages are obvious. First, the actual previous state ${\bf
x}^0$ is likely to be a better first guess for the modified initial
state $ {\bf x}^{*}$  of the predictor than it is for the final
state of the corrector step. Second, by modifying the non-linear
function and consequently modifying the Jacobian equation, the PC
preconditioner modifies the spectral properties of the Jacobian
matrix in the same way as  preconditioners applied directly to the
Jacobian equation. Indeed, as shown below the PC preconditioner
gives the same type of speed-up of the Krylov convergence without
requiring to formulate an actual preconditioning of the Krylov
solver.

We use a non-lnear diffusion problem and the standard driven cavity
flow problem as benchmarks to demonstrate the preformance and the
reliability of the PC preconditioning method.

\section{The Preconditioned Newton-Krylov Approach}

Most discretization schemes can be expressed as a set of difference
equations for a set of unknowns $\bf x$ representing the unknown
fields on a spatial  grid. Once the time is discretized, the state
vector $\bf x$ is computed at a sequence of discrete time levels. We
label the initial state of a time step (corresponding to the final
time of the previous time step) as ${\bf x}^0$ and the final time as
${\bf x}^1$.

When the time discretization scheme is fully implicit, the most
general two-level scheme can be formulated as a non-linear
relationship between ${\bf x}^0$ and ${\bf x}^1$:
\begin{equation}
f({\bf x}^0,{\bf x}^1)=0 \label{fi}
\end{equation}
where the vector function $f$ depends both on the initial and the
final states. The implicit nature of the scheme resides in the fact
that the function $f$ is a function of the new time level, requiring
the solution of a set of non-linear (if the function $f$ is
non-linear) coupled equations. As noted above this can be
accomplished with the NK method~\cite{kelley}. The method is based
on solving the Jacobian equation obtained linearizing the difference
eq.~(\ref{fi}) around the current available estimate ${\bf x}_{k}^1$
of the solution in the Newton iteration:
\begin{equation}
f({\bf x}^0,{\bf x}_k^1)+J\delta {\bf x} =0\label{jacobian}
\end{equation}
where $J=\partial f/\partial {\bf x}^1$ is the Jacobian matrix and
$\delta{\bf x}$ is the correction leading to the new estimation by
the Newton iteration: ${\bf x}_{k+1}^1={\bf x}_k^1+\delta {\bf x}$.

The solution of eq.~(\ref{jacobian}) is conducted with a Krylov
solver. The Jacobian matrix is approximated by a difference:
\begin{equation}
J\delta {\bf x} =\frac{f({\bf x}^0,{\bf x}_k^1+\epsilon \delta {\bf
x})-f({\bf x}^0,{\bf x}_k^1)}{\epsilon}\label{jacobian2}
\end{equation}
with $\epsilon$ chosen according to the machine
precision~\cite{kelley}.  Here we use the inexact Newton
method~\cite{saad-nlkrylov}, based on relaxing the convergence
criterion on the solution of the Jacobian equation when the Newton
equation is still far from convergence and progressively tightening
it as the Newton iterations close in on the solution. The specific
algorithm used here follows closely the implementation in the
textbook by Kelley~\cite{kelley}. For the Krylov solver we use GMRES~\cite{saad-gmres}
since the Jacobian matrix can be non symmetric.

While the pure NK method works in giving a solution, the number of
Krylov iterations required for each Newton step to solve
eq.~(\ref{jacobian}) can be staggering. In particular, as the grid
is refined and the size of the unknown vector ${\bf x}^1$ is
increased the number of Krylov iterations tends to increase. This is
the reason why a preconditioner is needed.

A remarkable feature of the NK method is its ability to be regarded
as a black box, communicating with the rest of the code only via the
evaluation of the non-linear function that summarize the discretized
partial differential equations. The NK black box provides as an
output a succession of guesses for the solution ${\bf x}^1_k$ and
requires as an input a residual coming from the function evaluation
${\bf r}_k$. At convergence, the residual is reduced to a prescribed
tolerance:
\begin{equation}
{\bf r}_k<\eta_a+\eta_r {\bf r}_0 \label{convergence}
\end{equation}
where the absolute tolerance $\eta_a$ and the relative tolerance
$\eta_r$ can be chosen by the user.

Most of the usual preconditioning techniques open the black box and
fiddle with the Krylov solver by enveloping a preconditioner around
the Krylov solver for the Jacobian equation to improve its
performance. This is accomplished in a number of very successful
methods that lead to nearly ideal performance~\cite{review}. The
approach is perfectly suited to new codes that can easily be
designed to implement the most effective preconditioners.

However, sometimes the need arises for retrofitting existing
semi-implicit codes. The goal in that case, is to use the existing
code as a preconditioner for a fully implicit approach. The new
fully implicit approach provides the non-linear function evaluation
for NK. The old code provides the preconditioner. In the usual
approach, the black box of the NK need to be opened and the old code
need to be modified to provide a representation for the Jacobian of
the new fully implicit method. A major part of the modification is
the fact that preconditioners act on Krylov vectors that are not
physical quantities but rather their change from Newton iteration to
Newton iteration as in eq.~(\ref{jacobian}).

Pre-existing codes operate on full fields, with their boundary
conditions, not on variations. The change to accommodate this need
can be considerable. We propose here a way to formulate
preconditioning techniques that does not require to operate on
variations but can operate directly on the fields themselves and
leaves the NK black box closed allowing the user to simply deploy
existing NK tools.

To arrive at the method, we remind that preconditioning can be
regarded in two ways. First, one can view preconditioning as a
modification of the Krylov step only. We focus here on the so-called
right preconditioning approach~\cite{kelley}. In that case, the
Jacobian problem is reformulated as:
\begin{equation}
f({\bf x}^0,{\bf x}_k^1)+ J P^{-1}P \delta {\bf x}
=0\label{jacobian_prec1}
\end{equation}
where the preconditioning matrix is chosen so that $JP^{-1}$ is an
approximation to the identity matrix, as it is when $P$ approximates
$J$. This point of view opens the NK black box and fiddles with the
Jacobian equation.

The approach followed here is based on an alternative look at the
preconditioning step presented in the classic textbook by
Kelley~\cite{kelley}. The preconditioning step can be regarded as a
modification of the non-linear function itself. The advantage of
this second perspective is two-fold. First, the black box of NK need
not be opened and the modifications required for preconditioning can
be done directly on the non-linear function evaluation. Second, this
point of view leads the way to an approach that acts directly on the
full fields, and not on their variations. As mentioned above, this
feature is key in retrofitting existing codes.

\section{Predictor-Corrector Preconditioners}
 In the present study, a preconditioner is constructed by using
 the predictor-corrector method. The key idea lies on modifying the
 non linear function evaluation that provides the residual for the NK iteration.

The approach requires to design alongside the fully implicit scheme
in eq.~(\ref{fi}), a second semi-implicit method. We note that this
is typically no hardship as semi-implicit methods were developed and
widely used before the implicit methods became tractable. Using the
same notation, we can write the most general two-level semi-implicit
algorithm as:
\begin{equation}
A {\bf x}^1 + f_{SI}({\bf x}^0) =0 \label{si}
\end{equation}
where $A$ is a linear operator (matrix) and the function $f_{SI}$
depends only on the old state ${\bf x}^0$. The semi-implicit nature
of the scheme resides on the fact that the difference eq.~(\ref{si})
depends non-linearly on the (known) old state ${\bf x}^0$ but only
linearly on the new (unknown) state ${\bf x}^1$.

In the classic implementation of preconditioners~\cite{review}, the
equation for the semi-implicit scheme (\ref{si}) is rewritten in
terms of the modification $\delta {\bf x}$ in a given Newton
iteration:
\begin{equation}
A\delta {\bf x}={\bf r}_k
\end{equation}
where ${\bf r}_k=f({\bf x}_k^1,{\bf x}^0)$ is the residual of the
current Newton iteration. The matrix $A$ of the semi-implicit scheme
becomes the preconditioner matrix $P$ for the Jacobian matrix $J$ of
eq.~(\ref{jacobian}). The approach has been extremely successful in
terms of providing a robust and effective solution scheme. For
example in the case of incompressible flows, the number of Krylov
iteration has been shown~\cite{pernice,knoll} to be reduced
drastically and to become nearly independent of the grid size.

However, a substantial modification of existing codes follows from
the need to modify the GMRES solver to use the matrix $A$ as a
preconditioner, especially when the method is formulated in a
matrix-free form where the matrix $J$ and the matrix $A$ are not
explicitly computed and stored.

We propose a different approach. We consider the following
predictor-corrector algorithm:
\begin{equation}
\left \{
\begin{array}{l}
\displaystyle ({\rm P})\; A {\bf x}^1 + f_{SI}({\bf x}^*) =0 \\
\displaystyle  ({\rm C})\; {\bf r}=f({\bf x}^0,{\bf x}^1)
\end{array} \right.
\label{pc}
\end{equation}
The predictor step uses the semi-implicit scheme to predict the new
state ${\bf x}^1$ starting from a modification of the initial state
${\bf x}^*$. The corrector step computes the residual ${\bf r}$ for
the fully implicit scheme when ${\bf x}^1$ from the predictor step
is used.

We propose to use scheme (\ref{pc}) by using ${\bf x}^0$ as the
initial guess of ${\bf x}^*$ and using the NK method to find the
solution for ${\bf x}^*$ that makes the residual ${\bf r}$ of the
corrector equation vanish. Once ${\bf r}=0$ (within a set
tolerance), the fully implicit scheme is solved, but it is solved
not iterating directly for ${\bf x}^1$ but iterating for the ${\bf
x}^*$ that makes the predictor step predict the correct solution
${\bf x}^1$ of the corrector step.

Two points are worth noting.

First, we have modified the task of the NK iteration changing our
unknown variable from ${\bf x}^1$ to ${\bf x}^*$. This corresponds
to change the non-linear residual function that the Newton method
needs to solve. To analyze this point, we consider a first order
Taylor series expansion of the preconditioned non-linear function:
\begin{equation}
f({\bf x}^0,-A^{-1}f_{SI}({\bf x}^*_k+\delta {\bf x}^*))=f({\bf
x}^0,-A^{-1}f_{SI}({\bf x}^*_k))-\frac{\partial f}{\partial {\bf
x}^1}A^{-1}\frac{\partial f_{SI}}{\partial {\bf x}^*}\delta {\bf
x}^*
\end{equation}

We observe that by first order Taylor expanding the preconditioner
step a relationship can be determined between the change$\delta {\bf
x}^*$ and the corresponding change in the end state $\delta {\bf
x}$:
\begin{equation}
A{\bf x^1_k}+A\delta {\bf x} + f_{SI}({\bf x}^*_k)+\frac{\partial
f_{SI}}{\partial {\bf x}^*}\delta {\bf x}^*=0
\end{equation}
Recalling that at the $k$-th iteration the preconditioner equation
was satisfied, i.e. $A{\bf x^1_k} + f_{SI}({\bf x}^*_k)=0$, it
follows that
\begin{equation}
\frac{\partial f_{SI}}{\partial {\bf x}^*} \delta {\bf x^*}=-A
\delta {\bf x}  \label{taylor_precond}
\end{equation}

Using eq.(\ref{taylor_precond}) and recalling the definition
$J=\partial f/\partial {\bf x}^1$, we can formally rewrite the
Jacobian equation for the preconditioned step:
\begin{equation}
JA^{-1}A\delta {\bf x}={\bf r}_k
\end{equation}
The equivalence of our approach to preconditioning and the standard
approach is now manifest. To first order in the Taylor series
expansion,  the new approach  is identical to applying the
traditional preconditioners directly to the Jacobian
equation~\cite{kelley}. However, to higher order this might be a
better approach as it reduces the distance between the initial guess
(${\bf x}^0$) and the solution for ${\bf x}^*$. If the semi-implicit
method works properly, ${\bf x}^0$ is closer to the converged ${\bf
x}^*$ than to the final state ${\bf x}^1$.

Second, programming the PC preconditioner is easier. The NK solver
can be used as a black box, without any need to formally go into it
and modify the Jacobian eq.~(\ref{jacobian}) by adding a
preconditioner. The semi-implicit method can be used directly on the
actual states and not on their variation $\delta{\bf x}$ between two
subsequent Newton iterates. This latter operation is complex as
boundary conditions and source terms in equations need to be treated
differently.

The approach described above is ideally suited for refitting an
existing semi-implicit code by simply taking an off the shelf NK
solver and wrapping it around the semi-implicit method already
implemented. The only change being that in the semi-implicit scheme
the initial state ${\bf x}^0$ is replaced by the guess of ${\bf
x}^*$ provided by the NK solver. We have indeed proceeded in this
fashion by wrapping the standard NK solver provided in the classic
textbook by Kelley~\cite{kelley} around our previously written
semi-implicit solver for the examples considered below.

\section{Numerical Experiments}
To test the method developed above, we present below two classic benchmarks: time-dependent diffusion in presence of  a non-linear diffusion coefficient and relaxation of an incompressible flow driven in a cavity.
\subsection{Non-linear Diffusion}
The non-linear diffusion equation in non dimensional units and in 1D cartesian geometry is:
\begin{equation}
\frac{\partial \phi}{\partial t}= \frac{\partial }{\partial x} \left
( D(\phi) \frac{\partial \phi}{\partial x} \right )
\end{equation}
where $\phi$ is the quantity being evolved. The diffusion
coefficient is chosen as $D(\phi)= \alpha_0 +\alpha_1 \phi$ with
$\alpha_0=\alpha_1/10$ and $\alpha_1=1$.

 The initial condition is $\phi=x \sin (x/L)/L$ and the
boundary conditions are $\phi(0)=0$ and $\phi(L)=0$ with $L=4$.

The method described above is implemented using a fully implicit
Crank-Nicolson scheme for the corrector step and a semi-implicit
scheme based on lagging the diffusion coefficient as predictor. In
space, the second order operator is discretized with centered
differencing.

The residual evaluation for the fully implicit (corrector) step is
\begin{equation}
r_i= \frac{\phi_i^1-\phi_i^0}{\Delta t} -
\frac{D_{i+1/2}^{1/2}}{\Delta x^2}\left(
\phi^{1/2}_{i+1}-\phi^{1/2}_{i}\right)+\frac{D_{i-1/2}^{1/2}}{\Delta
x^2}\left( \phi^{1/2}_i-\phi^{1/2}_{i-1}\right)
\label{nld:corrector}
\end{equation}
where $\phi^{1/2}_i=(\phi^{1}_i+\phi^{0}_i)/2$ and
$D^{1/2}_{i+1/2}=D((\phi^{1/2}_{i+1}+\phi^{1/2}_i)/2)$.

For the  predictor step we use:
\begin{equation}
\frac{\phi_i^1-\phi_i^*}{\Delta t} = \frac{D_{i+1/2}^{0}}{\Delta
x^2}\left(
\phi^{1/2*}_{i+1}-\phi^{1/2*}_{i}\right)-\frac{D_{i-1/2}^{0}}{\Delta
x^2}\left( \phi^{1/2*}_i-\phi^{1/2*}_{i-1}\right)
\label{nld:predictor}
\end{equation}
where $\phi^{1/2*}_i=(\phi^{1}_i+\phi^{*}_i)/2$ and the diffusion
coefficient is lagged using the actual value of the old time level
$\phi^0$. As prescribed in the method described above, the predictor
is modified by introducing the fictitious initial value $\phi^*_i$
that is the actual unknown solved for by the NK method.

The procedure is as follows:
\begin{enumerate}
\item The NK method
provides the guess of $\phi^*_i$ (the initial guess being the old
state $\phi^0$).
\item The predictor equation \ref{nld:predictor} is solved (using a simple tridiagonal
solver) for the advanced value of the unknown $\phi^1_i$
\item The residual of the corrector equation \ref{nld:corrector} is
computed and fed back to the NK solver to compute the new guess of
$\phi^*_i$.
\end{enumerate}

To solve the non-linear system we developed a program that
used the NK solver downloaded by the web site relative to the
textbook by Kelley \cite{kelley}. For the convergence
test, eq.~(\ref{convergence}), we used  $\eta_r=10^{-5}$,
$\eta_a=10^{-5}$ and we deployed the Eisenstat-Walker inexact
convergence criterion for the solution of the Jacobian
equation~\cite{eisenstat-walker}. The Krylov solver used is GMRES~\cite{saad-gmres}.

\begin{figure}
\centering
\includegraphics[width=8cm]{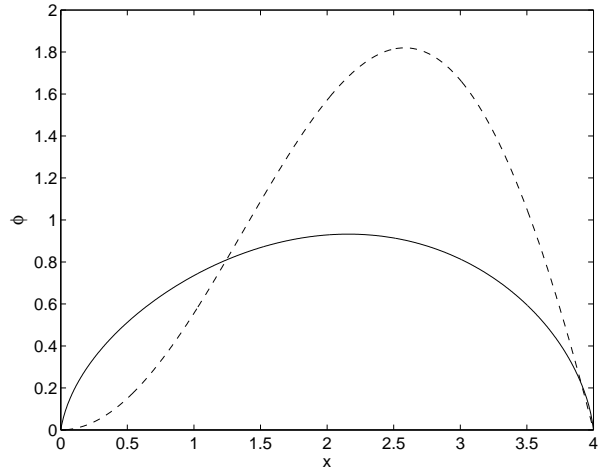}
\caption{Fiducial solution (obtained on a 800 cells grid) for the
non linear diffusion problem. The initial state is the dotted line,
the final state (t=1) is the solid line.} \label{nld_ref}
\end{figure}

 The problem does not have an analytical solution and
we report for reference the initial and final states (at $t=1.0$) in
the most refined simulation considered (800 cells),
Fig.~\ref{nld_ref}.

We have compared the efficiency of the NK solver with and without
the PC preconditioner described above. We change the number of cells
but hold the time step at $\Delta t=10^{-1}$, continuing the
simulation until the final time $t=1.0$. Table.~\ref{table_nld}
reports the number of Newton and Krylov iterations for different
grid sizes. Three trends emerge clearly. First, the number of Krylov
iterations is reduced drastically by the proconditioner in all
cases, even on the coarsest grid. Second, the number of Krylov
iterations in the unpreconditioned case scales in a very unfavorable
way with the number of cells increasing as the grid is refined,
compounding the increased cost of each iterations. Instead, with the
preconditioner the number of iterations remain essentially constant
on all grids, resulting in a virtually ideal scaling. Finally, even
with the present tests conducted with a straight-forward non-optimized
code, the actual CPU time of the preconditioned runs is always
smaller, resulting in a factor of 7 saving in the most refined grid.
We point out that this is rather remarkable in a simple 1D test. The
savings to be expected in 2D would be compounded by the
dimensionality.

\begin{table}
\caption{Number of Newton and Krylov (GMRES)  iterations and CPU
time (in seconds) for different uniform grids, with and without
preconditining. Tests conducted on a Windows XP operating system
with 1GB ram memory and Pentium M 1.4GHz.}
\begin{tabular}{||c||c|c|c||c|c|c||}
\hline
Grid &\multicolumn{3}{|c||}{Preconditioned} & \multicolumn{3}{|c||}{Un-Preconditioned} \\
 \hline N & Newton & Krylov & CPU Time & Newton & Krylov & CPU Time\\
\hline
100 & 2.82 & 3.18 & 0.19& 3.09 & 15.74 &0.32\\
200 & 3.00 & 3.64 & 0.28& 3.09 & 29.21 &0.67\\
400 & 3.00 & 3.82 & 0.43& 3.09 & 53.80 &1.75\\
800 & 3.00 & 4.67 & 0.93& 3.82 & 109.40 &6.42\\ \hline
\end{tabular}
\label{table_nld}
\end{table}
\subsection{Driven Cavity Flow}

\begin{figure}
\centering
\includegraphics[width=8cm]{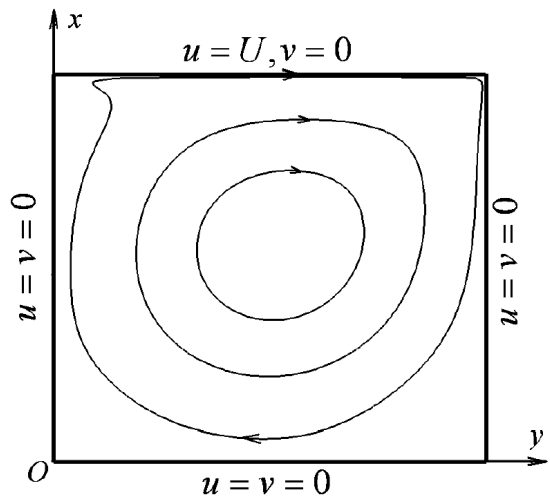}
\caption{Standard driven cavity flow and the velocity boundary
conditions} \label{cavity}
\end{figure}
 Next we consider the standard driven 2-D incompressible cavity
flow~\cite{chia}. The geometry and the velocity boundary conditions are shown in
Fig.~\ref{cavity}. The following non-dimensional variables are
introduced:
 \begin{equation}
(u,v)=\frac{(\hat{u},\hat{v})}{U}\,,\;(x,y)=\frac{(\hat{x},\hat{y})}{L}
\label{normal}
 \end{equation}
where the hatted variables represent the dimensional variables. The
scales are the cavity width $L$ and the upper boundary velocity $U$.
Time is normalized accordingly. The governing equation, in term of
the vorticity $\omega$, and the stream function $\psi$, can be
expressed as:
\begin{equation}
\frac{\partial^2\psi }{\partial x^2}+\frac{\partial^2\psi }{\partial
y^2}=-\omega \label{stream}
\end{equation}
\begin{equation}
\frac{\partial \omega}{\partial t}+u\frac{\partial \omega }{\partial
x}+v\frac{\partial \omega }{\partial y}=\frac{1}{{\rm
Re}}\left(\frac{\partial^2\omega }{\partial
x^2}+\frac{\partial^2\omega }{\partial y^2}\right) \label{vorticity}
\end{equation}
where $u=\partial \psi/\partial y$, $v=-\partial \psi/\partial x$,
$\omega=\partial v/\partial x-\partial u/\partial y$ , and ${\rm
Re}=UL/\nu$ is the Reynolds number based on the viscosity $\nu$.
Dirichlet  boundary conditions are applied for the stream function
and the boundary conditions for vorticity is determined by the
physical boundary conditions on the velocity~\cite{roache}. For
example, at the left wall $\omega =\partial v/\partial
x=-\partial^2\psi/\partial y^2$. We can obtain expressions for
$\omega$ at other walls in an analogous manner.

Eqs.~(\ref{stream}) and (\ref{vorticity}) are discretized using the
centered difference scheme in space. Two types of discretization are
considered in time, the semi-implicit Euler scheme (where the
velocity in eq.(\ref{vorticity}) is taken at the initial time level
of the time step) and the fully implicit Euler scheme.

We test the PC preconditioner approach using a fully implicit
backward Euler scheme for corrector:
\begin{equation}
r_{ij}=\frac{\omega^1_{ij}-\omega^0_{ij}}{\Delta t}+{\bf u}^1_{ij}
\cdot {\mathbf \Gamma}_{ij}(\omega^1)-\frac{1}{\rm Re}
\Delta_{ij}(\omega^1)
\end{equation}
where the discretized gradient (Laplacian) operator is indicated as
${\mathbf \Gamma}_{ij}(\omega^1)$ ($\Delta_{ij}(\omega^1)$) and is
evaluated using the new state $\omega^1$ in a fully implicit form.
The velocity is computed also from the new state using the stream
function elliptic equations in discretized form:
\begin{equation}
\Delta_{ij}(\psi^1)=-\omega^1_{ij}
\end{equation}

For preconditioning we use a semi-implicit scheme linearized by
computing the velocity using the previous Newton iteration
\begin{equation}
\frac{\omega^1_{ij}-\omega^*_{ij}}{\Delta t}+{\bf u}^{1(k)}_{ij}
\cdot {\mathbf \Gamma}_{ij}(\omega^1)-\frac{1}{\rm Re}
\Delta_{ij}(\omega^1)
\end{equation}
where the velocity is computed as:
\begin{equation}
\Delta_{ij}(\psi^1_k)=-\omega^{1(k)}_{ij}
\end{equation}
using the known vorticity from the previous Newton iteration. We
remark that using the old velocity rather than the previous guess
from the Newton iteration results in much poorer performances.

The code for the present test has been developed in Java using the
prescriptions of the textbook by Kelley~\cite{kelley} as reference.
The details of the implementation are identical as in the text
above. We remark incidentally that Java is a suitable scientific
computing language providing in its latest releases a competitive
computing performance when compared with C++, C or even
Fortran~\cite{markidis}. The resulting method is completely
matrix-free as only matrix-vector products, rather than details of
the matrix itself are needed. This circumstance greatly simplifies
the application of the method to complex problems.

 For reference, we present
results for a case with a mesh of 129$\times$129 cells. The classic
cavity flow solution is computed starting from a stagnant flow,
allowing the boundary conditions to drive the cavity to a steady
state. The flow condition at steady state is shown in Fig.
\ref{vorticity}. The figure is generated using the same contour
lines used in the reference benchmark solution presented by Chia et
al.~\cite{chia}. We compared  our solution with the published
reference benchmark obtaining complete agreement.

\begin{figure}
\centering
\begin{tabular}{cc}
\includegraphics[width=6cm]{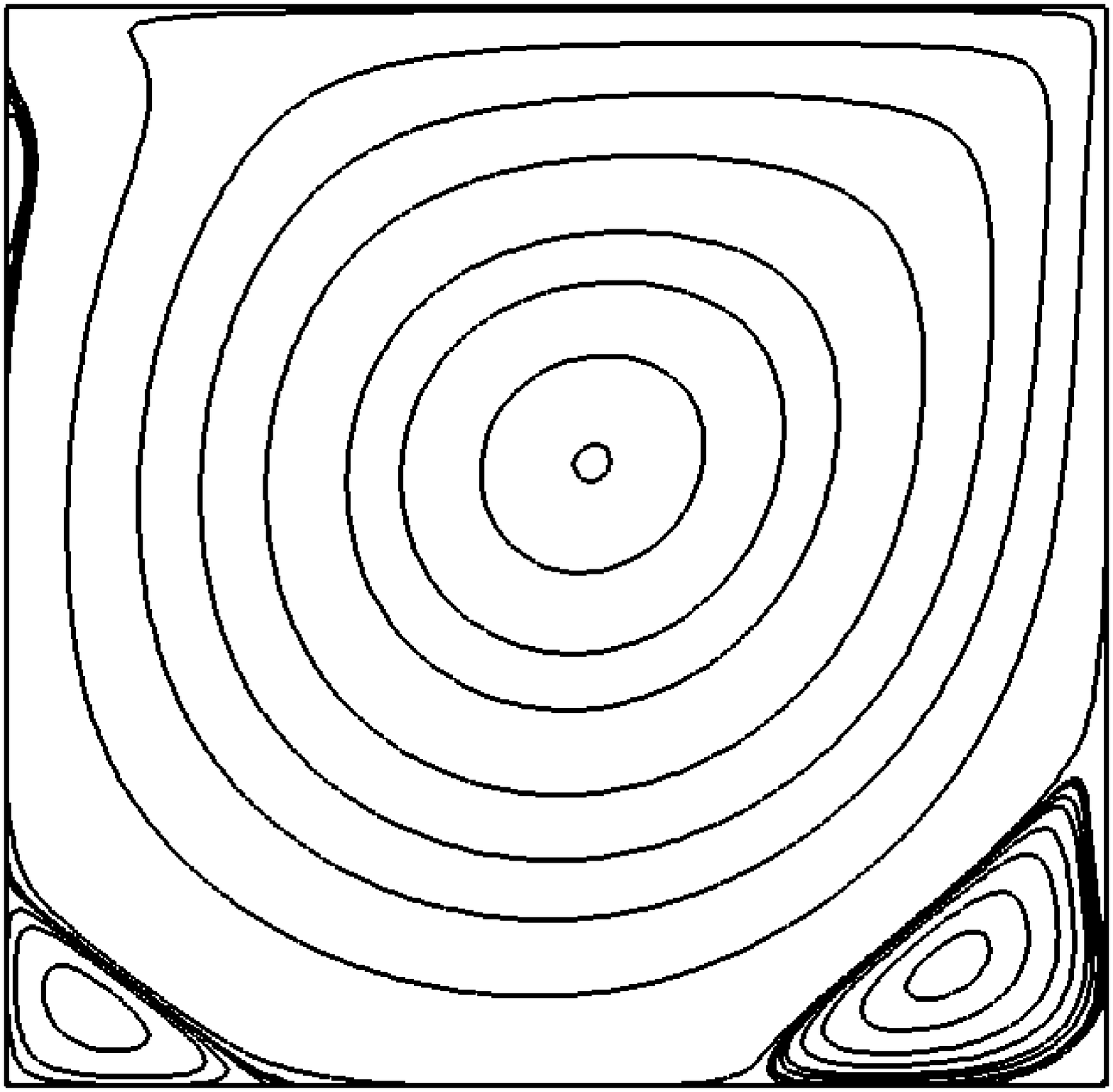} &\includegraphics[width=6cm]{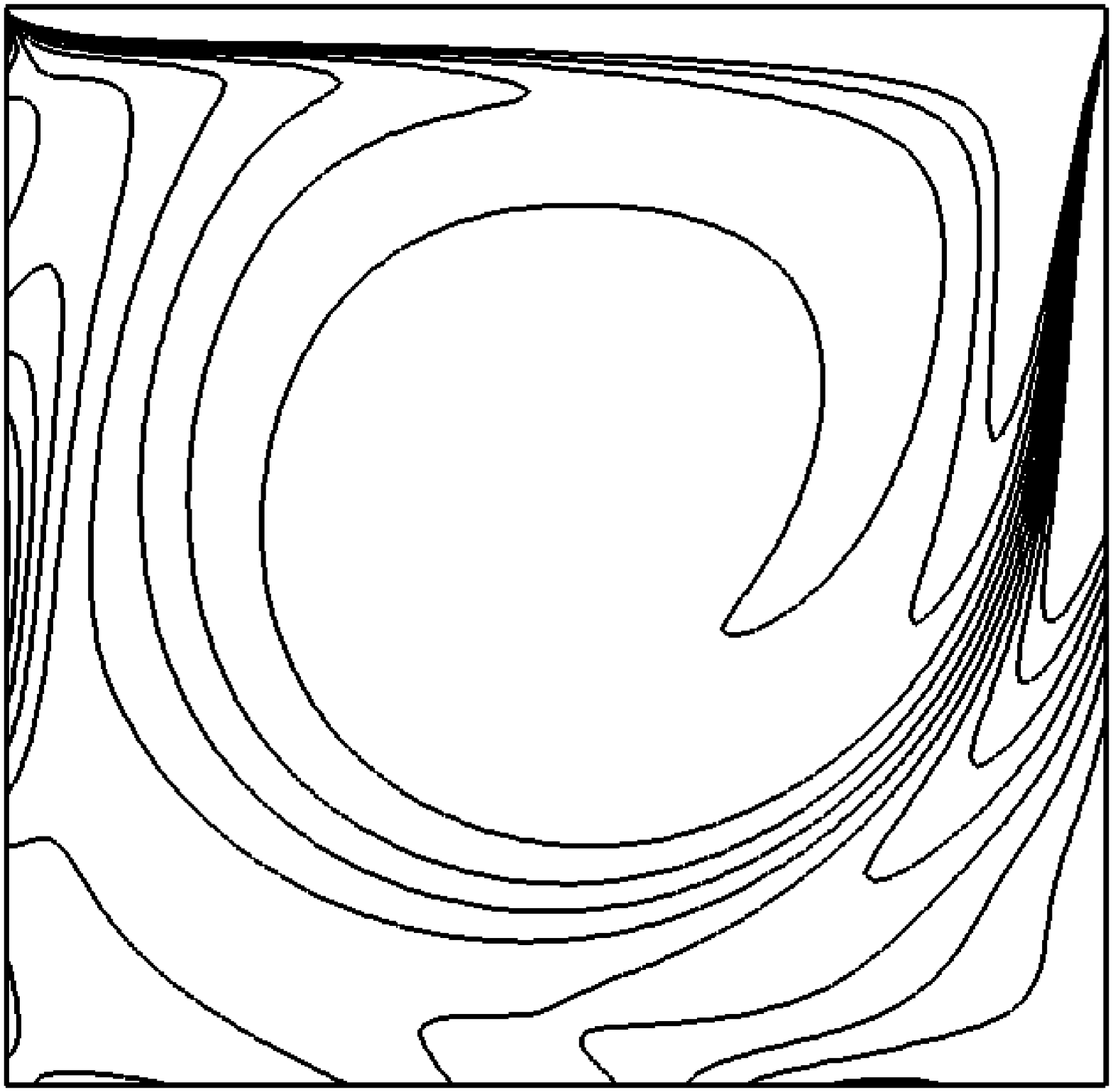}
\end{tabular}
\label{vorticity_stream}
 \caption{(a) Contours of stream function, (b) Contours of vorticity.
 Flow structure at  steady state for Re=1000}
\end{figure}

We have compared the efficiency of the NK solver with and without
the PC preconditioner described above. For the case without
preconditioner, the number of Newton and GMRES iterations is
reported in Table \ref{table_flow}. In the preconditioned case,
GMRES is never actually called thanks to the nearly perfect
performance of the preconditioner that removes the need for multiple
GMRES iterations. Table \ref{table_flow} reports the number of
Newton iteration, corresponding also to the number of calls to the
preconditioner, for this case. Total CPU times for the
preconditioned and unpreconditioned case is also reported.

As one can readily see, the number of GMRES iterations increases
without bounds in the unpreconditioned case, resulting in a
corresponding unbounded increase in computational costs. In the last
case in the table, the implicit case did not converge within the
maximum allotted number of iterations allowed.

In contrast, the preconditioned case, requiring only 1 iteration of
the preconditioner per Newton iterations, keeps the cost under
control and converging in all cases considered. In the most refined
cases, the gain exceed ten-fold, not mentioning the case where the
unpreconditioned run failed.

\begin{table}
\caption{Number of Krylov (GMRES) iterations, calls of
preconditioner and CPU time (in seconds) for different uniform grids
and time step, of the fully implicit case and preconditioned
semi-implicit cases. Tests conducted on a Windows XP operating
system with 1GB ram memory and Pentium M 1.4GHz.}
\begin{tabular}{||c||c||c||c|c||c|c||}
\hline
Grid &\ time step &\multicolumn{3}{|c||}{Unpreconditioned} & \multicolumn{2}{|c||}{Preconditioned} \\
 \hline N & $\Delta t$ & Newton & Krylov & CPU Time & Newton & CPU Time\\
\hline
10 & 0.01 & 1.01 & 79.2 & 4& 2.00 & 1\\
10 & 0.05 & 1.70 & 191.2 & 11& 3.01 & 2\\
10 & 0.1 & 2.00 & 239.6 & 15& 3.37 & 3\\
20 & 0.01 & 2.00 & 159.6 & 68& 3.00 & 11\\
20 & 0.05 & 2.01 & 275.2 & 120& 3.81 & 20\\
20 & 0.1 & 2.01 & 339.6 & 135& 4.19 & 37\\
40 & 0.01 & 2.00 & 235.2 & 829& 3.03 & 79\\
40 & 0.05 & 2.03 & 849.6 & 3122& 4.28 & 308\\
40 & 0.1 & 2.26 & 2732.4 & 9667& 4.96 & 661\\
60 & 0.01 & 2.01 & 287.6 & 2962& 3.24 & 192\\
60 & 0.05 & 2.23 & 1798.8 & 15465& 4.73 & 1042\\
60 & 0.1 & -- & -- & --& 5.31 & 1811\\ \hline
\end{tabular}
\label{table_flow}
\end{table}

\section{Conclusions}

 We presented a new implementation for preconditioning techniques based on using semi-implicit
 schemes to precondition fully implicit schemes. The fundamental new
 idea is to use the semi-implicit scheme as predictor and the fully
 implicit scheme as corrector, iterating the NK method on a
 modification of the old state used as initial state for the predictor rather than
 iterating on the final state of the corrector step as is typically
 done.

 There is one primary advantage to the new implementation.
Simplicity. The approach has
 been developed specifically with the goal in mind of reusing
 off-the-shelf existing semi-implicit methods and codes without
 requiring any modifications. In particular, the new implementation
 does not require to formulate the preconditioning step in terms of
 changes from a reference step (the previous Newton iteration). This
 latter requirement of previous approaches typically requires to make
 substantial modifications to existing code and n particular to
 boundary conditions. This requirement is completely eliminated, we
 require only one change: the initial state for the value of the
 state variables at each time step is no longer given by the old
 state but by the NK iteration.

 Most researchers and institutions have invested human efforts and capital
 in developing extremely sophisticated semi-implicit codes. Our
 approach allows to reutilize the invested effort without virtually
 any modification. An existing semi-implicit code can be built upon by supplying
 a new non-linear function evaluation for the new fully implicit scheme and a  NK solver.
 Off-the-shelf NK solvers are available easily from  freely available libraries such as
 TRILINOS~\cite{trilinos} or
 PETSc~\cite{petsc-manual} or can be easily implemented following the
 recipes of excellent textbooks~\cite{kelley} (the latter is the
 approach we followed).

\section*{Acknowledgments}

 Stimulating discussions with Luis Chac\'on and
Dana Knoll  are gratefully acknowledged. This research is supported
by the United States Department of Energy, under contract
W-7405-ENG-36.

\section*{References}

\bibliographystyle{unsrt}
\bibliography{precond}

\end{document}